\documentclass[graybox, envcountchap, authoryear, natbib]{svmult}

\usepackage{graphicx}        

\def\aap{{\em Astron. Astrophys. }}
\def\aj{{\em Astron.~J. }}
\def\apj{{\em Ap.~J. }}
\def\araa{{\em Annu. Rev. Astron. Astrophys. }}
\def\apjl{{\em Ap.~J.~Lett. }}
\def\apjs{{\em Ap.~J.~Suppl. }}

\def\mnras{{\em MNRAS }}

\def\pasa{{\em Publ. Astron. Soc. Aust. }}

\def\kms   {\ km s$^{-1}$}

\def\Msun  {${\rm M}_\odot$}
\def\deg   {$^\circ$}
\newcommand{\hi}{H\,{\footnotesize I}}

\begin{document}

\title*{An Introduction to Gas Accretion onto Galaxies}
\author{Mary E. Putman}
\institute{Mary E. Putman \at Department of Astronomy, Columbia University, New York, NY 10027, USA, \email{mputman@astro.columbia.edu}}
%
%
\maketitle
\abstract{
Evidence for gas accretion onto galaxies can be found throughout the universe.  In this chapter, I summarize the direct and indirect signatures of this process and discuss the primary sources.   The evidence for gas accretion includes the star formation rates and metallicities of galaxies, the evolution of the cold gas content of the universe with time, numerous indirect indicators for individual galaxies, and a few direct detections of inflow.  The primary sources of gas accretion are the intergalactic medium, satellite gas and feedback material.  There is support for each of these sources from observations and simulations, but the methods with which the fuel ultimately settles in to form stars remain murky. }

\section{Introduction}
\label{sec:1}

The idea of gas accretion onto galaxies first came $\sim50$ years ago. In the 1960s and 70s, observations of high velocity hydrogen clouds were made and it was proposed they represent infalling Galactic fuel \citep{muller63,hulsbosch68,dieter71,oort70}.  This infall was soon understood to have consequences on the Milky Way's star formation and distribution of stellar metallicities \citep{larson72a, larson72b,vandenbergh62}.
Since the original detection of Galactic cold hydrogen halo clouds, observations have been made of halo gas in a variety of phases for galaxies throughout the universe.  The observations have made it clear that there are abundant baryons surrounding galaxies and that some of these baryons will accrete and fuel future star formation.  

There are numerous observations of galaxies and the intergalactic medium (IGM) that infer gas accretion is needed throughout cosmic time.  There are galaxies at all redshifts observable with star formation rates that indicate they will run out of fuel within a few Gyrs without replenishment.  The metallicities of their stars also suggest galaxy evolution models without accretion will not work.   At the level of the census of baryons in the universe, the decrease in the mass density of cold hydrogen with time does not closely track the steady increase in the mass density of stars, and this requires the ionized IGM to cool and accrete onto galaxies.   The first section of this introduction provides an overview of this type of indirect observational evidence for gas accretion.

Theoretically, ongoing gas accretion is required to produce realistic galaxies and it largely occurs through the accretion of the intergalactic medium and satellites.  Feedback from galaxies is also a key component of galaxy formation models and the interplay between accretion and feedback is important to understand.   There is some observational support for the IGM, satellite gas, and feedback material all being sources of future star formation fuel.  The second section of this chapter discusses these sources of gas accretion and the different modes with which the gas may ultimately reach the star-forming core of a galaxy.

Direct evidence for gas accretion, as in the kinematic signature of gas falling directly onto the stellar component of a galaxy, is relatively rare and the third section of this chapter summarizes the direct observational evidence currently available.   Some of the numerous additional observational claims for gas accretion onto individual systems are also discussed.
I conclude this chapter with a brief summary and thoughts on directions for the future.

\section{The Need for Accretion through Cosmic Time}
\label{sec:2}

The need for ongoing gas accretion is evident in observations of galaxies and the IGM at all redshifts.   
In this section, the broader indirect pieces of observational evidence for gas accretion are discussed.   This includes the star formation rates of populations of galaxies, the state of the baryons in the universe, and the metal enrichment history of galaxies.

Star formation rates that will exhaust a galaxy's gas supply on a relatively rapid timescale are commonly derived from observations.  At high redshift ($z>1$), there are measured gas depletion times of less than a Gyr \citep[e.g.][]{genzel10, daddi10, tacconi13}.   The high redshift observations are limited to those galaxies with accessible gas and star formation tracers \citep{shapley11}, but this result is found for a wide variety of tracers.  At lower redshift, where there is a more complete census of the gas content and star formation rate of galaxies, the measured gas depletion times are longer, but still typically less than a few Gyrs \citep{kennicutt12, bigiel11, leroy13, schiminovich10}.    The primary exception is the gas-rich dwarf galaxies that can have depletion times closer to a Hubble time \citep{vanzee01,hunt15, huang12}.   There is evidence that galaxies are undergoing a more efficient mode of star formation at higher redshift, leading to the shorter depletion times and the need for a greater rate of accretion at early times \citep{scoville16, santini14}.  This is also evident in the evolution of the specific star formation rate (SFR/M$_*$) that gradually decreases towards lower redshifts at a given stellar mass \citep{madau14, karim11}.   The accretion process may be key to regulating a galaxy's SFR and without it the depletion times at all redshifts suggest a large fraction of galaxies are not far from being red and dead.  

\begin{figure}[t]
\begin{center}
\includegraphics[scale=.47]{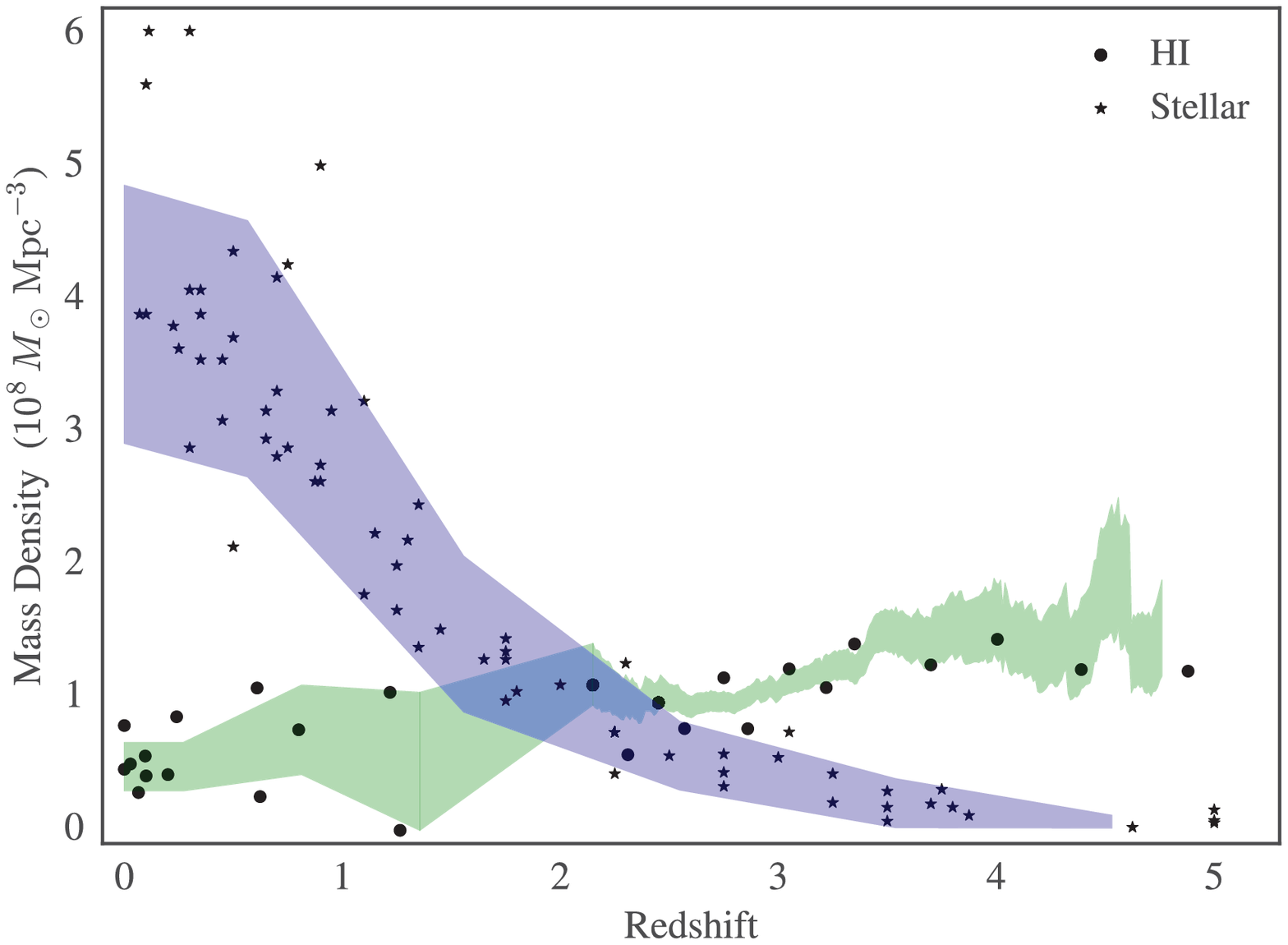}
\includegraphics[scale=.47]{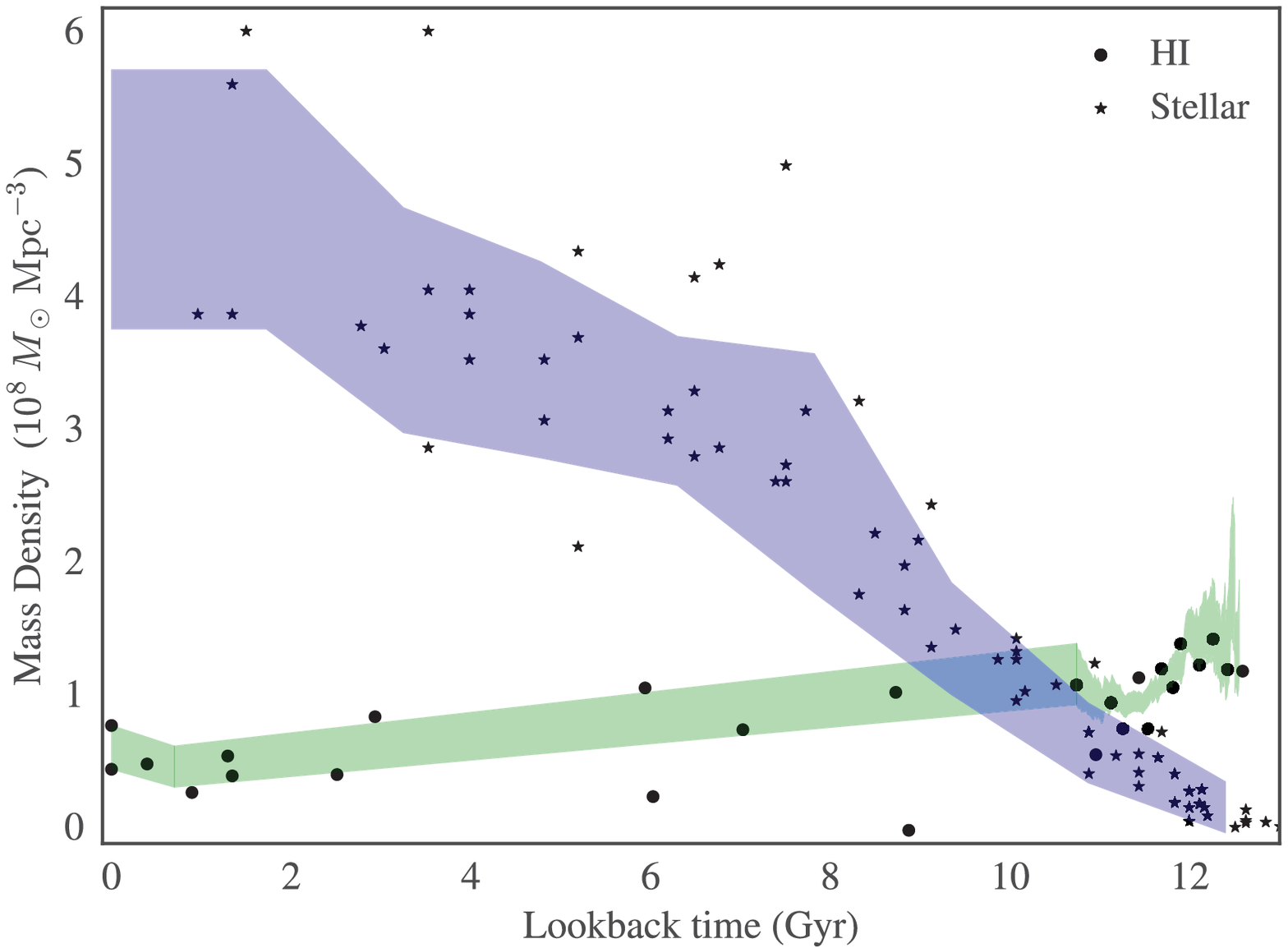}
\end{center}
\caption{A comparison of the evolution of the stellar mass density (stars/purple) and atomic gas (HI) mass density (points/green) with redshift (top) and lookback time (bottom).   The HI data are from the compilation by \cite{neeleman16} and \cite{sanchez16}, and the stellar mass densities are taken from the \cite{madau14} compilation.  The HI measurements are largely based on damped Lyman-$\alpha$ absorption measurements except at $z\sim 0$.  The purple and green shades represent the running median and scatter of the data with the shaded HI incorporating the $z>1.6$ results of \cite{sanchez16}.  }
\label{fig:rhohist}       
\end{figure}

Beyond individual galaxies, the cold gas content of the universe as a whole should decrease as more stars are formed.   If it does not correspondingly decrease there must be continuous cooling of the ionized gas in the IGM and halos that harbor the majority of the baryons in the universe \citep{shull12,bregman07}.   This can be investigated with a comparison of the mass density of atomic hydrogen to the mass density of stars through cosmic time, as shown in Figure~\ref{fig:rhohist}.  This figure shows that the HI mass density has an overall decrease from $z=3$ to $z=0$ ($\sim11$ Gyr ago to today), but this evolution is mild compared to the increase in stellar mass density with time.   Where the decrease in HI mass density specifically happens depends on the measurements adopted \citep{rao06,lah07,prochaska05}, but the values at $z=3$ derived from damped Lyman-$\alpha$ absorbers (DLAs) are clearly higher than the HI emission measurements at $z=0$ \citep{zwaan05,hoppmann15}.    A recent study by \cite{neeleman16} at $z\sim0.6$ ($\sim 6$ Gyr ago) is consistent with a gradual decline from $z=2$ to today.  This measurement was achieved with a blind DLA survey, an improvement over the higher point at approximately this redshift/time that is based on Mg II surveys \citep[see][]{prochaska09}. 
In any case, the evolution of the HI does not appear to closely follow the evolution of the stellar mass density or star formation rate density of the universe \citep{madau14, putman09a,hopkins08}.    Naively a direct correlation would be expected if there is no accretion and cooling of new HI gas.  

There is clearly more work to be done to understand the evolution of baryons across time.  For the mass density of atomic hydrogen, we are currently limited to using absorption line studies at every redshift but $z=0$, where HI emission measurements are available.  This will improve with HI emission surveys that can reach $z=0.5$ ($\sim5$ Gyr ago) in progress with the JVLA \citep{fernandez16}, and planned with SKA precursor telescopes  \citep[ASKAP; ][]{duffy12}.  The MeerKAT survey LADUMA is designed to detect HI in emission out to at least $z=1$ (over half the age of the universe), and this will significantly add to our knowledge of the evolution of the HI mass density.  
CO surveys with ALMA are also an important component to our future understanding of the evolution of the cold gas content of the universe.  The CO has a direct correlation with star formation \citep[i.e., a short consumption time; ][]{bigiel11,leroy08}, and correspondingly the molecular (traced by CO) to stellar mass ratio already shows indications of a clear decline with redshift \citep{carilli13,bauermeister13}.  The molecular gas depletion rate does still require continuous replenishment via gas inflow \citep{bauermeister10}, but it is not as directly apparent as with the HI evolution.  

The final piece of indirect observational evidence for accretion discussed in this section is the metallicity distribution of the stars in galaxies.  In the local universe, the metallicity distribution of the long-lived stars support galaxy evolution models with a continuous inflow of relatively low metallicity gas \citep{chiappini09,fenner03,larson72b}.   Closed box galaxy evolution models produce a wider distribution of stellar metallicities than is observed and this has traditionally been referred to as the G-dwarf problem.   This evidence for accretion is strongest in the local universe where the metallicity of individual long-lived stars can be measured \citep{holmberg07,kirby13,grillmair96}, but it is also consistent with the metallicities derived from the integrated stellar light observations of galaxies \citep[e.g.][]{henry99,bressan94,stott14,gallazzi05}.  The metallicity distribution of planetary nebulae has also been used as evidence for gas accretion \citep[e.g.][]{magrini07}.  These metallicity results are for a variety of galaxy types and the fact that the accreting gas needs to be relatively low metallicity is considered support for the IGM being a major source of accretion.

\section{Expected Modes of Accretion}
\label{sec:2}

\begin{figure}[!ht]
\begin{center}
\includegraphics[height=1.5in]{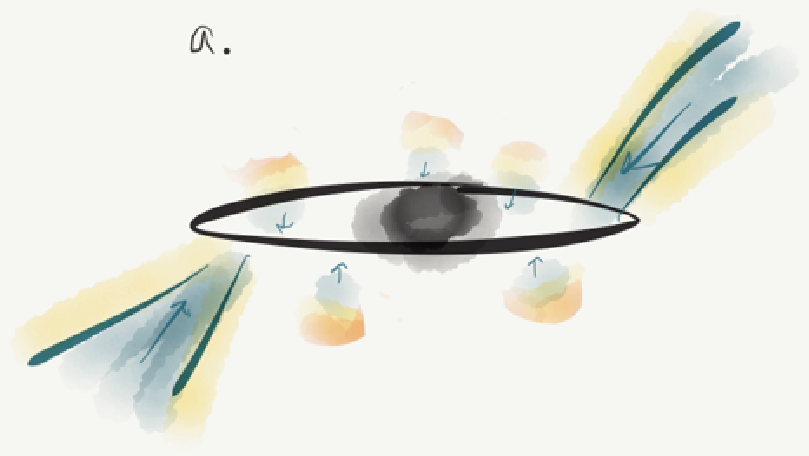}
\includegraphics[height=1.5in]{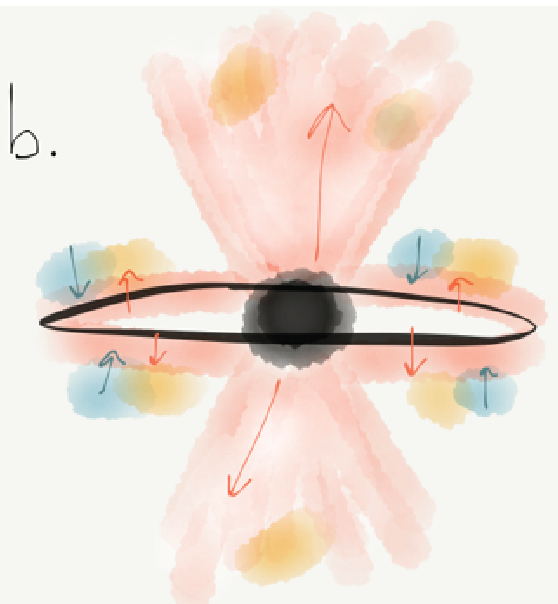}
\includegraphics[height=1.5in]{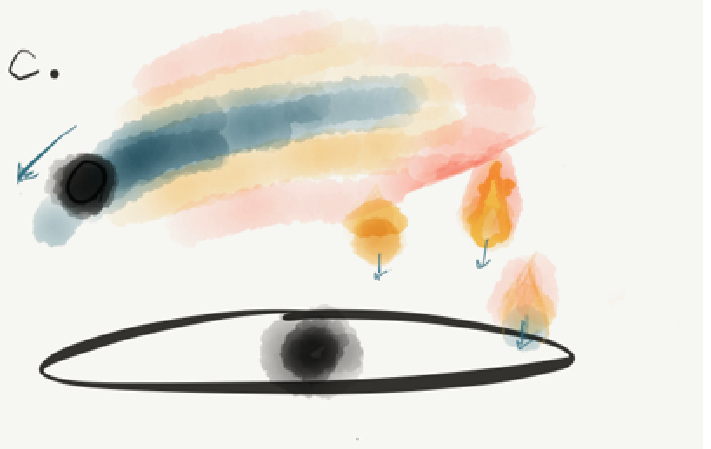}
\caption{A representation of the three expected sources of accretion with red indicating hot gas and blue cooler gas.  
a) (top-left) Accretion from the IGM along filaments where the outer parts are heated and the inner parts are able to cool. The hot IGM-originated halo gas cooling near the disk as it mixes with denser gas is also indicated. 
b) (top-right) Feedback material can accrete as part of a fountain flow close to the disk with hot gas from stellar feedback rising and then cooling and falling back down.  Gas from a central outflow will mix with existing halo density enhancements and this may also result in cool clumps that eventually accrete. 
c) (bottom) Satellites are stripped of their gas as they move through the diffuse halo medium and this gas will fall to the disk as warm clouds.   As the gas slows and mixes with denser feedback material it can potentially re-cool close to the disk.}
\label{fig:cart}       
\end{center}
\end{figure}

Gas accretion onto the stellar component of a galaxy can proceed in several ways and from multiple sources.   Most of the gas in the halos of galaxies is ionized, and since the ionized gas mass in a galaxy's disk is smaller than the mass in cold gas \citep[$<10^4$ K; e.g.,][]{ferriere01}, the halo gas must rapidly cool as it accretes.   The gas may come in as large cool clouds, dribble onto the disk from smaller warm clouds, enter preferentially at the edges of the galaxy, or the gas may accrete with a combination of these methods.  The dominant mode of accretion at the star forming component of a galaxy remains to be determined.  The interplay between enriched outflowing gas with gas coming in may be key in this process \citep{fraternali08,putman12,marinacci10,voit15}.

\begin{figure}[!ht]
\sidecaption[t]
\includegraphics[scale=.68, bb= 100 0 300 400]{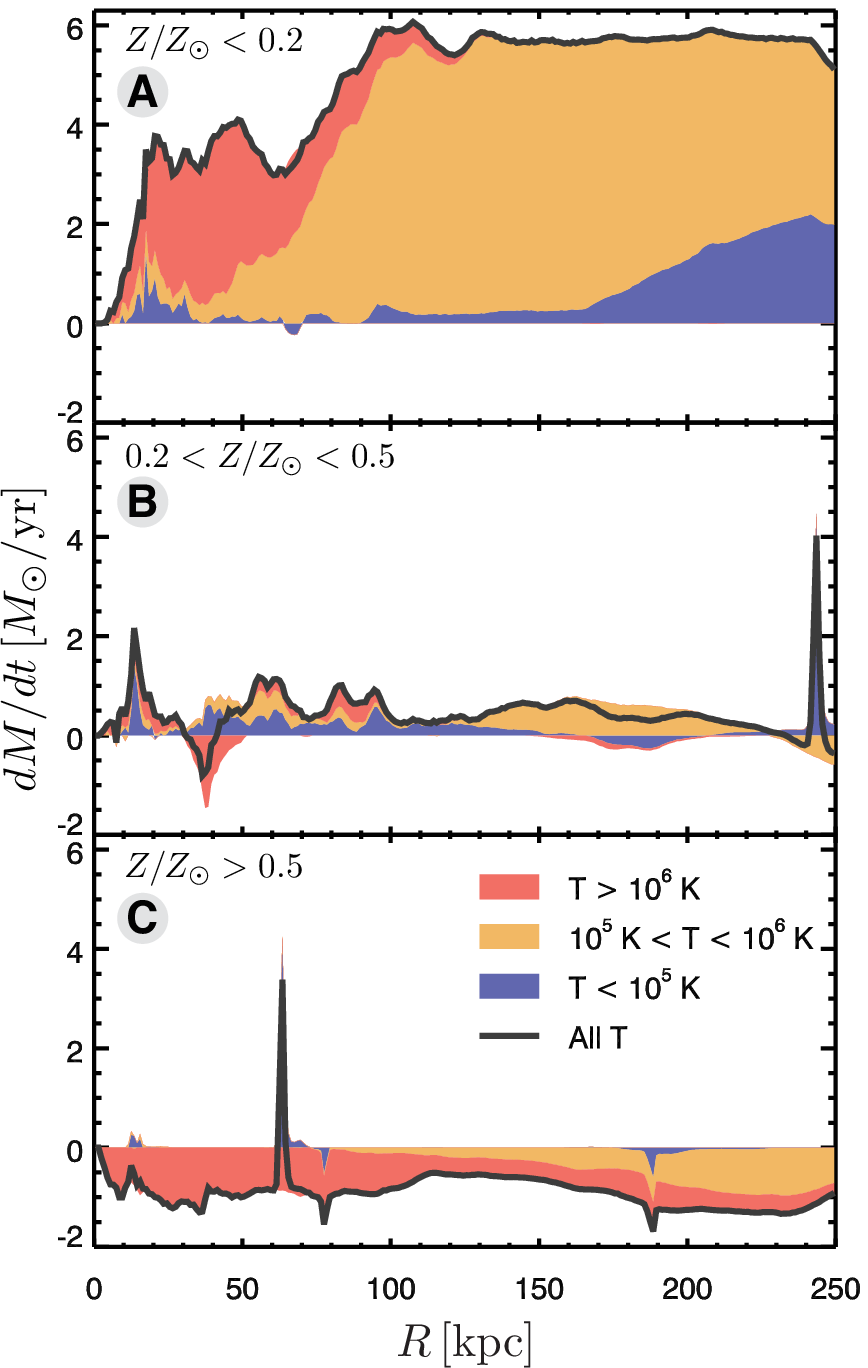}
\caption{The mass accretion rate of different temperature and metallicity gas at $z=0$ in an AMR simulation of a Milky Way mass galaxy \citep{joung12b}.  Most of the accreting gas is low metallicity material from the IGM as shown in panel A.   The highest metallicity gas (panel C) has a net outflow at all radii.  The change in temperature of the accreting gas is largely due to the heating of the inflowing gas as it interacts with existing halo gas and the cooling in the central regions of the filaments.}
\label{fig:sim}       
\end{figure}

The major sources of the accreting material are thought to be the IGM, satellites, and recycled feedback gas (see Figure~\ref{fig:cart}).   
Theoretically, all three of these sources are expected and observationally there is evidence for all three, although much of the evidence is indirect.  For instance, the continuous distribution of gas from a galaxy through its halo to the IGM, as found with absorption line experiments \citep{tumlinson13,prochaska11,penton02,wakker09,chen01}, is consistent with the IGM as an important fuel source, though not direct evidence of its accretion.  Theoretically, the inflowing filaments of IGM are expected to be the largest source of ongoing accretion for a galaxy \citep{joung12b,keres05,brooks09}.   Depending on the mass of the galaxy halo, some percentage of the inflowing IGM is heated to high temperatures in the simulations.  Figure~\ref{fig:sim} shows the state of the accreting case for a L$_*$ galaxy and how the filaments of low metallicity IGM (top panel) are partially heated as they move through the halo and also cool in the central regions as they approach the disk.   How the gas is able to ultimately cool to below $10^4$ K and feed the star formation in the disk may be related to density enhancements in the filaments and the mixing with satellite and feedback material \citep{joung12b,fraternali08}.
The simulations also find that much of the ongoing IGM accretion occurs towards the edges of the galaxy to avoid the dominant feedback from the central regions \citep{stewart11,fernandez12}.

\begin{figure}[!ht]
\begin{center}
\includegraphics[scale=.3]{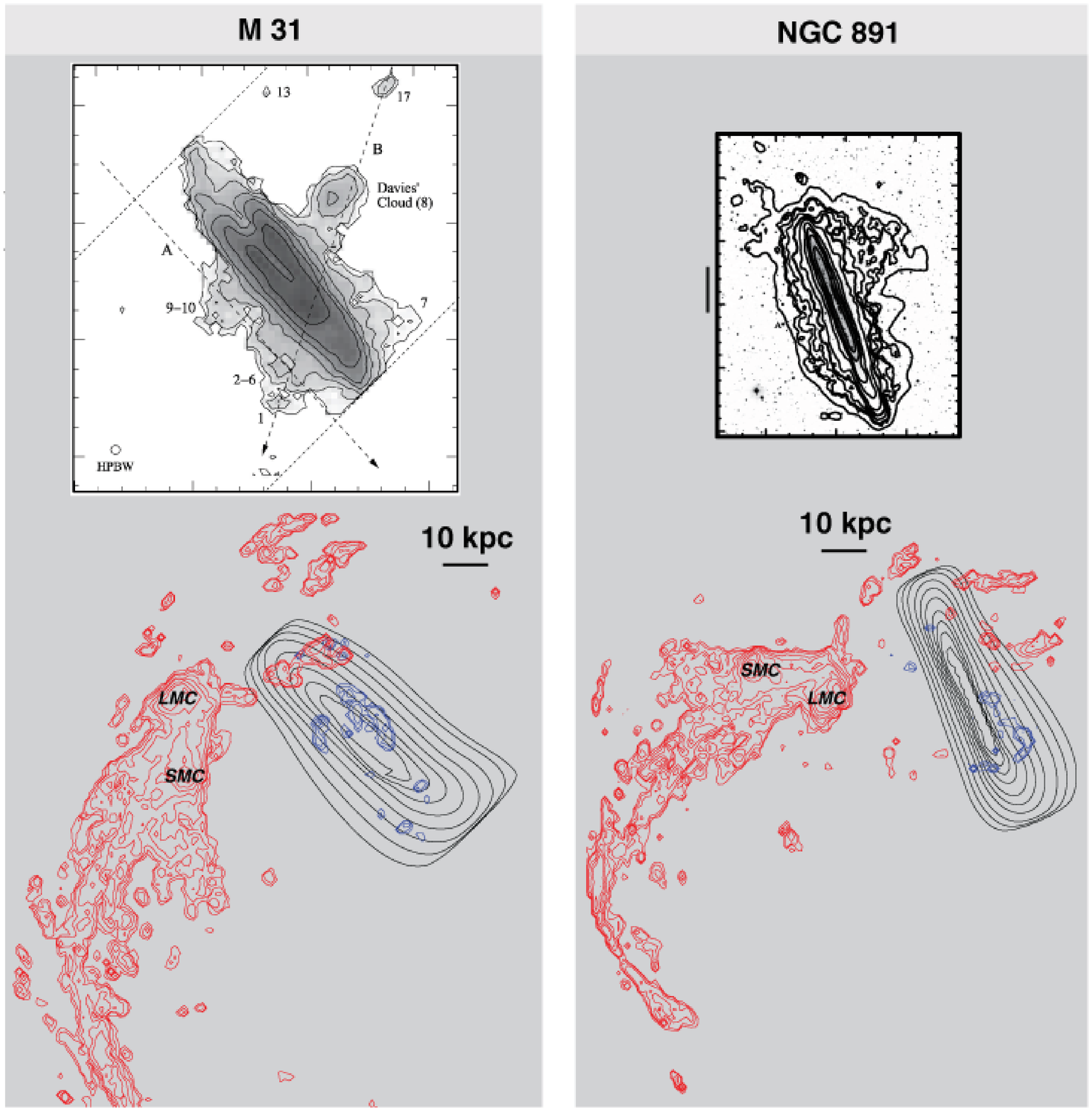}
\caption{Evidence for accretion from \hi\ observations for the Milky Way (bottom), M31 \citep[top left;][]{westmeier07}, and NGC~891 \citep[top right;][]{oosterloo07}.  The Milky Way observations are shown from an external point of view with the viewing angle and distance to M31 (bottom left) and NGC~891 (bottom right).  The accreting Milky Way \hi\ halo gas at $<10$ kpc (blue contours) is difficult to discern from the disk gas (black contours) with an external view.   The red contours show the accretion of satellite \hi\ from the Magellanic System \citep{putman03}.  The origin of the extraplanar \hi\ for M31 and NGC~891 is unknown, but some of the larger \hi\ features are potentially linked to satellite accretion.   The extraplanar \hi\ shown is not a substantial amount of mass, but some of it may represent the cooling of the large reservoir of ionized halo gas.   This figure is from \cite{putman12} and more details can be found there.}
\label{fig:hi}    
\end{center}   
\end{figure}

It is clear that satellite gas is stripped within a galaxy halo as observations have captured this directly (e.g., Figure~\ref{fig:hi}), and satellites closer to galaxies typically do not have gas and are redder \citep{grcevich09, spekkens14, geha12}.  Ram pressure stripping by the CGM of the host galaxy is thought to be the dominant stripping mechanism, but other forces can be important, for instance when satellites come in as an interacting pair \citep{pearson16,marasco16}.  The gas is largely heated when it is ram pressure stripped \citep{tepper15, gatto13,fox14}, and (again) it is not completely clear how it ultimately cools to feed the galaxy's star formation.  It may sink to the disk as density enhancements in the halo and ultimately cool closer to the disk as it slows and encounters a denser surrounding medium \citep{heitsch09,joung12b,blandhawthorn07}.   The numerous small satellites found in the Local Group and predicted by simulations do not provide a significant amount of gas, but larger satellites, such as the Magellanic Clouds for the Milky Way, can provide gigayears worth of star formation fuel to a galaxy.

There are numerous observed kinematic signatures of feedback mechanisms putting gas into the halos of galaxies \citep{rubin14,shapley03,weiner09,chen10,heckman00}.   While simulations require many of the metals created by a galaxy are ejected from the system (see panel c of Figure~\ref{fig:sim}), there is a large mass of metals detected in galaxy halos that remains bound \citep{tumlinson11,werk14}.  The results therefore indicate there is abundant future star formation fuel that has already cycled through the galaxy.    As discussed at the end of \S2, the metallicities of the stars in galaxies indicate feedback should not dominate as the fuel source.  Mixing feedback material with IGM and satellite gas is key to balance this out.  This is consistent with the results of simulations that can produce the large amount of detected ions in galaxy halos while remaining consistent with the observed mass-metallicity relation for galaxies \citep{oppenheimer16,muratov16}.

\section{Direct Observational Evidence for Accretion}
\label{sec:3}

Direct unambiguous kinematic evidence of gas falling onto a galaxy is relatively rare.   The simple approximation of the mass accretion rate is $\dot{\rm M}$ (\Msun$/$yr) = Mv$/$z, where M is the mass of the accreting material, v is its velocity, and z is the height the material is falling from.   Each of these parameters usually has significant uncertainties in the observations depending on the gas phase probed and the geometry of the system.  In all cases we are only capable of measuring one component of the gas velocity, and accretion can be more accurately assessed when the gas is known to be close to the disk and unmeasured tangential velocities cannot easily dominate.

The Milky Way is an example of gas observed at 1-15 kpc above the disk that is clearly infalling.   The actual rate of accretion depends on the 3D motions of the gas and the full extent of the accreting layer, but the rates calculated are 0.1-0.4 \Msun$/$yr for the coldest gas \citep{putman12}, and closer to 1 \Msun$/$yr when the ionized gas is included \citep{lehner11}.  Most of the gas thought to be in the halo of the Milky Way has unknown distances.  The only halo gas known to be at large radii is that associated with the Magellanic System \citep{putman03,fox14}.   It is difficult to say when the gas of Magellanic origin will accrete as it is likely to have a large tangential velocity component and will slow and be heated as it falls.  An extended, diffuse halo medium is inferred to exist for the Milky Way from the nature of the Magellanic System and stripped satellites \citep{salem15,emerick16,grcevich09}.  The motion of this diffuse halo medium is largely unknown, and much of it is thought to be hot and not easily observed.  There is at least a consistency between the radial velocities of the ions that probe the warm-hot halo gas and the simulated extended halo medium represented in Figure~\ref{fig:sim} \citep{zheng15}.  

There is limited direct evidence for accretion beyond the Milky Way.   Absorption line experiments that use background QSOs do not know the location and motion of the gas relative to the galaxy's stars.  Experiments that use objects within the galaxy itself do not have the uncertainty of the gas being on the near or far side and the corresponding ambiguity of the velocity potentially representing inflow or outflow.   The vast majority of the observations using the galaxy itself show significant outflows, with only a few examples of detected inflow.  \cite{rubin12} and \cite{martin12} detected Mg II and Fe II absorption for $100+$ star forming galaxies at $z=0.4-1.3$ and found only 4-6\% show cool gas inflow with velocities $< 200$ \kms.  The low detection rate may be related to the covering fraction of the cold gas inflow, the low velocities of the inflow relative to their velocity resolution, and/or significant inflow could be an intermittent process.  \cite{rubin12} found all but one of the galaxies with inflow have an inclination $> 60$\deg; and this may have helped to differentiate the inflowing material from the ubiquitous outflows.  Many other claims of inflow that use the galaxy itself are tenuous given the velocity resolution of the observations and the difficulty in separating the inflow component from the galaxy in the spectrum \citep{sato09, giavalisco11}.   

\begin{figure}[t]
\includegraphics[width=4.6in]{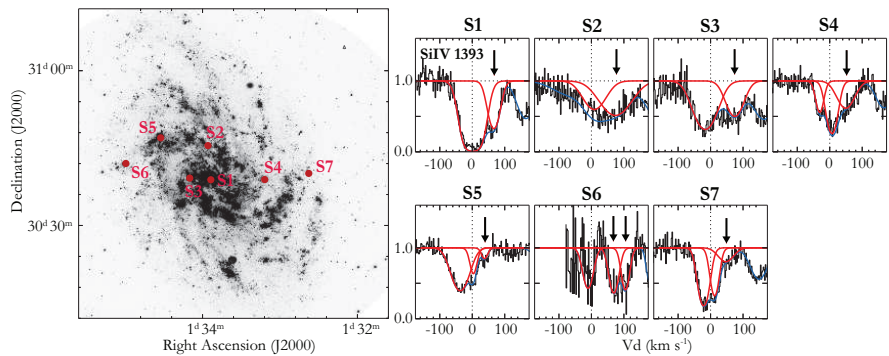}
\caption{The direct detection of accretion for the Local Group dwarf spiral galaxy M33 (Zheng et al. 2016).   The left is a UV image of M33's star formation from {\it GALEX} with the stars targeted with the Cosmic Origins Spectrograph on {\it HST} labeled as S1-S7.   The right shows the Si IV ($\lambda$1393) absorption lines detected for each sightline.  The 0 \kms~dotted line is the observed systemic velocity of the rotating \hi\ disk, the red represents the fits to the individual lines and the blue is the composite fit.   The arrows indicate gas that is inflowing with respect to M33's disk.}
\label{fig:m33}       
\end{figure}

One of the strongest cases for the direct detection of gas accretion beyond the Milky Way is for the small spiral galaxy M33 in the Local Group.  Zheng et al. (2016) used UV-bright stars in M33's disk as background probes and found the kinematic signature of inflow across the star forming disk in the Si IV absorption lines (Figure~\ref{fig:m33}).  An accreting layer of gas at the disk-halo interface is the most consistent model with the distribution and velocities of the inflowing gas.  A layer close to the disk is consonant with the difficulty in detecting the inflow in other systems.   The accretion rate obtained ($\sim2.9$ \Msun$/$yr) is relatively large for this small galaxy, and   may be further evidence for the infall of fuel being intermittent in nature.

With the numerous indirect methods of detecting gas accretion it is difficult to provide a complete census of the results.   This paragraph gives an overview of some of these methods.   As mentioned previously, absorption line experiments that use distant background probes and model the likely location of the gas relative to the galaxy are often used to claim accretion \citep{bouche16,bowen16}.  The actual location of the gas is not known, but cases where the absorbing gas is close in position-velocity space to the galaxy are more likely to be capturing the accretion process.  
  When particularly low metallicity gas is detected in a galaxy halo it is often claimed to be the accretion of the IGM filaments mentioned in \S3 \citep[e.g.,][]{lehner13,cooper15,crighton13}.  It would be difficult to explain this gas as anything else, but since the exact location and kinematics of the gas is unknown, it may or may not be accreting.  It is also difficult to separate an IGM origin from satellite material at low redshift using metallicity alone \citep{muratov16}.
Filamentary extensions of Lyman-$\alpha$ emission have been taken to be the detection of accretion at high redshift.  In cases when multiple datasets are combined, the evidence is particularly strong \citep[e.g.][]{rauch16,fumagalli16,martin16}.
Finally, at low redshift, extensions of \hi\ emission have been published as evidence for accretion \citep[e.g.][top panels of Figure~\ref{fig:hi}]{kreckel12,putman09a,sancisi08}.   The gas detected will certainly eventually accrete onto the nearby galaxy, but the origin of the gas and the direction the gas is currently moving is usually uncertain.


\section{Summary}

As outlined in this chapter and throughout this book, it is clear that gas accretion onto galaxies is occurring.  We now know that the accreting cold hydrogen clouds originally found for the Milky Way are a small component of a process found throughout the universe.  There are clear kinematic signatures of infalling gas for multiple galaxies and numerous other observations of individual systems that are consistent with gas accretion (\S4).  Beyond the evidence in individual systems, galaxies throughout time have star formation rates and metallicities that require ongoing accretion and the evolution of the \hi\ mass density with time suggests it needs a constant source of replenishment (\S2).  The three main sources of accretion seen in simulations (IGM, satellites, feedback material) are consistent with what is found in observations of galaxy halos (\S3).   

Though we have made tremendous progress from when the idea of gas accretion originated, there remain many open questions as to exactly how gas accretion proceeds and how frequently.   In particular, how the abundant ionized baryons within galaxy halos ultimately become star formation fuel is unclear, as is whether the infall occurrence is preferentially linked to satellite accretion or a feedback cycle.
Future observations and simulations should further investigate how accretion and feedback are consistent with each other.  In simulations the two mechanisms often preferentially occur along different axes, but observations have yet to solidly confirm this.   It is also possible that accretion turns on when a galaxy is not in as dominant of a feedback stage.   We are well-poised with the future direction of observations and simulations to address these questions.  Sensitive, high velocity resolution absorption line experiments that use the galaxy itself combined with spatially and kinematically resolved observations of galaxies in emission will reveal additional clear cases of gas accretion and begin to place the origin of the gas.   Sophisticated, high resolution simulations that include realistic feedback prescriptions and gas mixing can be compared to observations and will serve as a useful guide to our understanding.

\begin{acknowledgement}
I thank J. Xavier Prochaska for his help making Figure 1 pretty, Marcel Neeleman for sharing his data, Yong Zheng for providing the M33 figure, Jim Putman for useful plotting advice, Dan and Elaine Putman for daycare support, and NSF support from grants AST-1410800 and AST-1312888.   Thanks to the University of Colorado for hosting me while I wrote this article, in particular conversations with Ben Oppenheimer and Julie Comerford.
\end{acknowledgement}
%

\end{document}